\documentclass[letterpaper]{article}
\usepackage{aaai}
\usepackage{times}
\usepackage{helvet}
\usepackage{courier}
\usepackage{color}
\usepackage{url}
\usepackage{graphicx}
\usepackage{subfigure}

\newcommand{\hide}[1]{}

\newcommand{\hdr}[1]{\vspace{1mm}\noindent{\bf{#1}}}
\def\spaceB{\vspace{-3mm}}
\def\spaceA{\vspace{-4mm}}
\newcommand{\denselistA}{\itemsep 0pt\topsep-5pt\partopsep-7pt }
\newcommand{\denselist}{\itemsep -1pt\topsep-5pt\partopsep-7pt }
\enlargethispage{\baselineskip} 

\begin{document}

\title{Governance in Social Media:\\ A case study of the Wikipedia promotion process}
\author{Jure Leskovec \\ Stanford University\\jure@cs.stanford.edu
\And Daniel Huttenlocher\\ Cornell University\\dph@cs.cornell.edu
\And Jon Kleinberg \\ Cornell University\\kleinber@cs.cornell.edu
}

\maketitle

\newcommand{\UnnumberedFootnote}[1]{{\def\thefootnote{}\footnote{#1}
\addtocounter{footnote}{-1}}}

\UnnumberedFootnote{
\hspace*{-18pt} Work supported in part by NSF grant IIS-0705774,
Yahoo Research alliance grant, Microsoft Research and IBM Faculty Award.
}

\begin{abstract}
Social media sites are often guided by a core group of committed users
engaged in various forms of {\em governance}.
A crucial aspect of this type of governance is {\em deliberation}, in which
such a group reaches decisions on issues of importance to the site.
Despite its crucial --- though subtle --- role in how a number
of prominent social media sites function,
there has been relatively little investigation of the 
deliberative aspects of social media governance.

Here we explore this issue, investigating a particular
deliberative process that is extensive, public, and recorded:
the promotion of Wikipedia admins, which is determined 
by elections that engage committed members of the Wikipedia community.
We find that the group decision-making at the heart of this process
exhibits several fundamental forms of {\em relative assessment}.
First we observe that
the chance that a voter will support a candidate is strongly
dependent on the relationship between characteristics of the 
voter and the candidate.  Second we investigate how
both individual voter decisions and overall election
outcomes can be based on models that take into account the
sequential, public nature of the voting.
\end{abstract}

\section{Introduction}
\label{sec:intro}
The overall behavior of a social media site is generally driven by
the collective activity of a large population, but in many cases these sites
are also guided
by a much smaller group of core participants who are
strongly committed to the success of the site. The guidance provided by such a
core group can take many forms, ranging from assignment of tasks in massive
open-source and crowdsourcing projects, to enforcement of explicitly
articulated norms and rules on a site like Wikipedia, to much more informal
types of on-line organizing, question-answering, and expertise location.

We think of all of these mechanisms as forms of {\em governance}, a process
that plays an important role in social media, despite the fact that it is
generally much more subtle --- and maintains a much lower profile --- than the
forms of political, legal, and corporate governance that we are familiar with
in the off-line world. Governance in social media involves both deliberation
(the reaching of decisions by a core group) and also enforcement (the
carrying-out of these decisions).  There has been interesting recent work
on governance in social media (see e.g. 
Beschastnikh et al. \shortcite{beschastnikh09wiki}),
but it remains a topic where there is much still to be understood ---
particularly on the issue of deliberation, since it
can be difficult to find records of the process by which decisions were
actually reached.

\begin{figure}[t]
    \centering
    \includegraphics[width=0.3\textwidth]{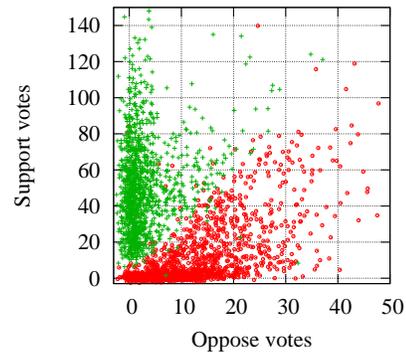}
    \spaceB
    \caption{
    A scatter plot of the
    number of supporting and opposing votes, and the outcome of each election.
    Elections leading to promotion to adminship are shown in green;
    elections that did not lead to promotion are shown in red.
    \label{fig:scatter}
    }
    \spaceA
\end{figure}

\hdr{The Present Work: Group Decision-Making and Wikipedia Promotion.} In this
paper we consider the deliberative aspects of social media governance, by
focusing on a setting where detailed traces of group decision-making by a
site's elite can be studied.

Our setting is the Wikipedia promotion process, in which users of Wikipedia can
be nominated to become {\em admins} --- a category of highly trusted user to
whom special administrative privileges are granted. The promotion process has a
clearly defined formal structure: the candidate for adminship submits a case
for promotion; there is then a period of discussion and deliberation by the
community; and this is followed by a vote. There are two important features of
this process that are worth noting. First, any Wikipedia user is allowed to
vote, not just users who have achieved admin status (although the contents and
results of the voting are interpreted by a special class of admins called
{\em bureaucrats} in order to reach a final decision).
Second, and crucial for our
research purposes, the discussion and voting is carried out completely in
public, and is recorded as part of Wikipedia, so that a transcript is
subsequently available. As a very simple illustration to give a sense for the
dataset, Figure~\ref{fig:scatter} shows a scatter-plot of the number of
positive and negative votes in each Wikipedia promotion election. We describe
the dataset in much more detail in a subsequent section.

The Wikipedia promotion process thus has the key ingredients we need: it is a
deliberative process carried out by core, committed members of a social-media
community; it has the goal of producing a single group decision; and it is
publicly recorded, making the analysis possible. It also serves as an instance
of a broad type of decision-making, familiar from the off-line world as well as
the on-line world, in which people are asked to offer evaluations of other
people.

Wikipedia promotion was studied recently by Burke and Kraut
\shortcite{burke08promotion}; their focus was on considering the process from
the perspective of candidates for adminship, using properties of the candidates
to develop statistical models capturing their likelihoods of promotion. In
contrast, because of our interest in the issue of deliberation, we study the
process from the perspective of the voters: we ask how voters evaluate
candidates, how a single voter behaves across many different elections, and how
voting unfolds over time as an election of a single candidate is carried out in
public.

\hdr{The Present Work: Main Results.} Our main findings can be viewed as
identifying ways in which a voter's evaluation of an admin candidate reflects
different types of {\em relative assessments} --- based on the relation of the
voter to the candidate, and to the (public) decisions of other voters.

We begin by analyzing how the relationship between characteristics of a voter
$V$ and a candidate $C$ affect $V$'s decision to vote positively or negatively.
We find that the probability $V$ will vote positively on $C$ is strongly
dependent on the relative values of several basic ``figures of merit'' for $C$
and $V$; these include which of $C$ or $V$ has a greater number of edits, and
which has a greater number of {\em barnstars}, awards given by other members of
Wikipedia~\cite{kriplean08barnstars}.\footnote{Here is the definition of a
barnstar from Wikipedia: {\em It is the custom to reward Wikipedia contributors
for hard work and due diligence by awarding them a barnstar. To give the award
to someone, just place the image on their talk page (or their awards page), and
say why you have given it to them. Wiki barnstars were introduced to Wikipedia
in December 2003. Since then, the concept has become ingrained in the Wikipedia
culture. These awards are part of the Kindness Campaign and are meant to
promote civility and WikiLove. They are a form of warm fuzzy: they are free to
give and they bring joy to the recipient.}} The extent to which $C$ and $V$
have interacted in the past also has a significant effect on the likelihood
that $V$ will vote positively on $C$. Overall, this analysis suggests that one
should think of the likelihood of a positive vote on a candidate not as a
function of just the candidate alone, but as a function of both the candidate
and the voter.

We then consider the relationship between a voter $V$'s decision and the public
decisions of other voters expressed earlier in the election. To make this
precise, we show how to compute a {\em response function} for $V$, giving the
probability that $V$ will vote positively as a function of the fraction of
preceding votes that were positive. A non-trivial number of Wikipedia editors
have each voted in several hundred elections, making reliable estimates of
their individual response functions possible. We find a striking level of
diversity in the response functions of these very frequent voters: some of them
are very stingy with their positive votes, while others are much freer. These
findings raise an intriguing possibility that transcends the particular
definition of response functions and addresses the broader issue of aggregation
in social data; it suggests that when we observe cumulative curves showing how
members of a population respond {\em in aggregate} to the behavior of others
(e.g. the types of social influence analyses found in Backstrom et al.
\shortcite{backstrom-kdd06}, Kossinets and Watts \shortcite{kossinets-email},
and Leskovec et al. \shortcite{leskovec-ec06}), it may be that these aggregate
functions are not typical of any particular individual, but instead represent
averages over populations that are highly heterogeneous.

Understanding the relationship among different voters' decisions involves the
consideration of how a single election's dynamics play out over time, as votes
are cast publicly in sequence. We explore this issue further, asking how these
dynamics affect the overall outcome of the election --- i.e., whether the group
decision is positive or negative. This is the setting of fundamental models for
information cascades in economic theory
\cite{banerjee-herding,bikhchandani-fads}, and as such it is interesting to see
how the dynamics reflected in the real data compare to the predictions of these
models. We find that the probability of an election's success depends heavily
on the outcomes of the first few votes --- primarily, one expects, because
these first few votes provide powerful evidence for the strength of the
promotion case.
However, with a few structured exceptions that we identify, we do not find
strong evidence that the {\em order} in which
a given set of initial positive and negative votes are interleaved has a 
significant effect on the outcome.
This forms an interesting contrast with
predictions of ``herding'' behavior, in which it
is argued that a few concurring votes at the
outset of a sequential voting process can induce subsequent conformity.

\section{Related Work}
\label{sec:related}
As discussed in the introduction, governance in social media includes both
deliberation and enforcement. The issue of enforcement, which is distinct from
our investigation here, has been the focus of a line of work in human-computer
interaction, addressing issues such as how the development
and application of norms can help
control deviant behavior in on-line communities (see e.g. Cosley et al.
\shortcite{cosley-norms} and the references therein).

In the study of deliberation, there have been recent investigations of on-line
settings in which public opinions are expressed sequentially, as they are in
Wikipedia promotion as well. Wu and Huberman \shortcite{wu08opinion} study the
sequences of reviews for a product on Amazon, and Danescu-Niculescu-Mizil et
al. \shortcite{mizil09opinions} study the question of how the helpfulness of
such reviews are evaluated by Amazon's user community. In a related vein,
Lerman \shortcite{lerman08digg} studies the patterns of voting for news stories
on Digg, identifying patterns that help predict whether a story will become
highly popular.

Finally, as noted in the introduction, Burke and Kraut
\shortcite{burke08promotion} have previously studied the Wikipedia promotion
process, but focusing on the characteristics of candidates rather than on
voters as we do here; they analyze textual features describing contributions
and user activity to build classifies for predicting adminship election
outcomes.

\section{Dataset description}
\label{sec:elecs}
In Wikipedia any user can be nominated for promotion to adminship. When a
candidate is considered for a promotion there is a public discussion and
vote.  Each vote is signed by the user who
produced it, and  votes are generally accompanied by some explanatory text
written by the voter.
After the election the record and all the votes are kept in the Wikipedia
archives. We collected data on all the elections in
the English part of Wikipedia
between September 17, 2004 and January 6, 2008, which gave us a set of 2,794
elections.

Votes can be cast in one of three categories: support, oppose and
neutral. In our dataset there is a total of 114,040 votes: 83,962 support,
23,118 oppose and 6,960 neutral. This yields a baseline probability of a
support vote of 0.784. Each vote can get discussed or commented on by
other users
and thus rich discussions can develop. Overall 7\% of support votes got
discussed, while 82\% of the oppose votes were commented on
or further discussed.

\begin{figure}[t]
    \centering
    \includegraphics[width=0.35\textwidth]{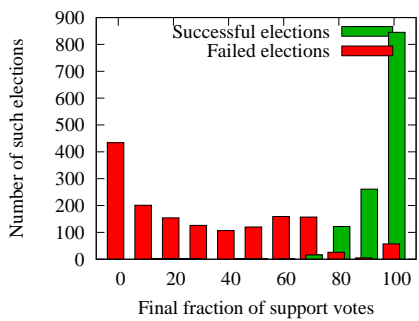}
    \spaceB
    \caption{
    Final fraction of support votes for elections that resulted in promotion
    and those that did not.
    \label{fig:sucfrac}
    }
    \spaceA
\end{figure}

When a candidate is considered for promotion, after about a week-long voting
period a Wikipedia bureaucrat in charge of overseing the election
decides whether the nomination for promotion was successful. The bureaucrat
makes this
decision based on a consensus of voting users. Overall 1,248 (44.6\%)
elections resulted in successful promotion. In successful elections (i.e.,
those that resulted in promotion) on average there were 52.2 support, 1.5
neutral and 3.1 oppose votes. On average successful elections concluded with
94.7\% of all votes supporting the promotion. For failed elections there were
on average 12.5 oppose, 3.3 neutral and 11.9 support votes, and these
elections
closed with an average of 31\% of the votes supporting the promotion.
Figure~\ref{fig:sucfrac} plots the histogram of final fraction of support votes
for successful and failed elections.

In principle, any registered Wikipedia user can cast a vote. However a very
small fraction them actually do so. 8,298 distinct users participated in
elections either as voters (7,499 users) and/or candidates (2,539 users).
Notice the number of candidates is smaller than the number of elections as a
single candidate can go up for election multiple times until he or she is
elected. Table~\ref{tab:users} gives statistics about the user population,
which we split into the following three disjoint classes:
\begin{itemize}
    \denselist
    \item \emph{Administrators} are users for which some election
        turned out successful.
    \item \emph{Unsuccessful candidates} are users who went up for
        promotion but the election(s) turned out unsuccessfully.
    \item \emph{Other users} are editors that only cast votes but were
        never considered for promotion to adminiship.
\end{itemize}

\begin{table}[t]
    \centering
    \small
    \begin{tabular}{l||r|r|r}
    User type & $N$ & $f_v$ & $p_s$ \\ \hline \hline
    Administrators             & 1,235 & 44\% & 0.794 \\
    Unsuccessful candidates  & 1,304 & 12\% & 0.748 \\
    Other users (voters)       & 5,759 & 44\% & 0.783 \\
    \end{tabular}
    \spaceB
    \caption{User statistics. $N$: number of
    users, $f_v$: fraction of votes casted, $p_s$: probability
    of a support vote.}
    \label{tab:users}
    \spaceA
\end{table}

For the analyses described in the remainder of the paper we discarded all
neutral votes and consider only elections with at least 10 votes. In a small
number of cases when a user changed her mind and recast the vote, we consider
the last vote cast by that user as the vote.

\section{Relative Merit of Candidates and Voters}
\label{sec:status}
We begin by considering properties of a voter $V$ and a candidate $C$ that
affect $V$'s decision on whether to vote positively or negatively on $C$.

To provide some context for this, recall that Burke and Kraut
\shortcite{burke08promotion} analyzed the success of a Wikipedia promotion
candidate $C$ based on characteristics exhibited by $C$. Their work leaves open
two qualitatively distinct possibilities. First, it may be that the probability
that $C$ receives a positive vote is a function primarily of $C$'s attributes
alone --- in other words, there are certain criteria for a person to become an
admin (i.e., a certain number of edits, a certain number of barnstars, and so
forth), and the decision of any voter is mainly an application of these common
criteria. Alternately, it may instead be that the probability of $C$ receiving
a positive vote depends in a significant way on the relationship between
characteristics of $C$ and characteristics of the voter $V$ who casts the vote.
In other words, we may want to model voter $V$ as performing a relative
assessment of $C$ through implicit (even if not overt) comparison to $V$'s own
merit. This latter possibility is particularly intriguing, given several recent
lines of work suggesting the importance of relative
comparisons between an individual and a peer group, in contrast to absolute
evaluations of merit \cite{burt-neighbor-networks-book,leskovec-chi10}.

We approach this question by computing univariate measures derived from
differences in merit between $C$ and $V$; this allows us to identify
one-dimensional relationships based on these differences. As we now show, there
is strong evidence that such measures of {\em relative merit} between $C$ and
$V$ play a significant role in the empirical probability that $V$ votes
positively on $C$.

\begin{figure}[t]
    \centering
    \begin{tabular}{c}
    \includegraphics[width=0.4\textwidth]{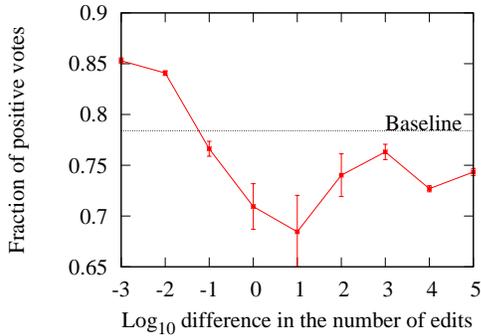}  \\
   (a) Positive vote fraction based on edit-count difference \\
    \includegraphics[width=0.4\textwidth]{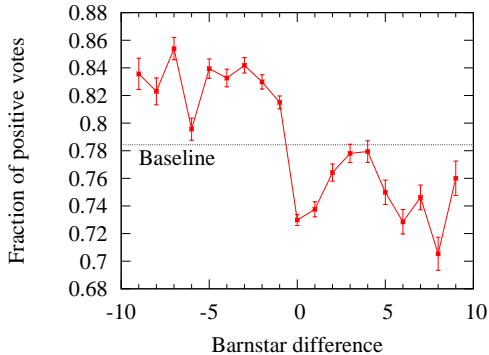}\\
   (b) Positive vote fraction based on barnstar difference
    \end{tabular}
    \spaceB
    \caption{
Probability of $V$ voting positively on $C$ given the difference in merit between $V$ and $C$.
(a) Difference in the number of edits between $V$ and $C$.
(b) Difference in the number of barnstars.
}
    \spaceA
    \label{fig:delta}
\end{figure}

\hdr{Relative Merit.} In our analysis of relative merit, we will be interested
in the {\em positive-vote fraction}: the overall fraction of positive votes
received by candidates from voters, restricted to different sub-populations of
the candidates and voters. We consider ways of evaluating a candidate $C$
relative to a voter $V$ based on different figures of merit.
We begin with the number of edits to articles, which can be taken as
a basic measure of the total activity (and hence, in some sense, contributions)
on Wikipedia.

In Figure~\ref{fig:delta}(a), we show the probability that a voter $V$ will
support a candidate $C$ as a function of the signed logarithm (${\rm sign}(x)
\cdot \log_{10}(|x|)$) of the difference in the number of edits they've each
made. (Thus a negative value means that $V$ has made fewer edits than $C$,
while a positive value means that $V$ has made more edits than $C$, and these
differences appear on the $x$-axis on a logarithmic scale.) We observe several
important features of this plot. First, it is significantly higher to the left
of $0$ (when candidate $C$ has more edits) than it is to the right of $0$ (when
voter $V$ has more edits). This is the most basic indication that the relative
merit of $V$ and $C$ is playing a role. Moreover, the effect on the
positive-vote fraction is significant over multiple orders of magnitude in the
difference of edit counts.

\hdr{Non-Monotonic Effects of Relative Merit.} There is a further striking
point to note about Figure~\ref{fig:delta}(a): not only is there a drop in the
positive-vote fraction as we move from negative log-differences to positive
ones, but there is also a ``rebound'' in which the positive vote fraction
climbs again once the log-difference exceeds $1$. (The error bars indicate that
this effect is significant.) This means that, in aggregate, voters are {\em
least likely} to support candidates who have edit counts that are approximately
the same as their own. Note that even though there is a rebound when $V$ has
higher edit count than $C$, the probability of $V$ voting positively is still
below the baseline.

In Figure~\ref{fig:delta}(b) we perform the same analysis for a different
relative figure of merit: the difference between the number of barnstars
received by the candidate $C$ and the voter $V$.  (Again, a negative difference
means that the candidate has more barnstars than the voter.) The shape of the
curve is surprisingly similar, given that the measure of merit is quite
different; again we see the drop from negative differences to positive ones,
and the same non-monotonicity around $0$. There is an additional interesting
feature in Figure~\ref{fig:delta}(b): the single biggest change in the
positive-vote fraction occurs when we move from negative barnstar differences
to non-negative barnstar differences.  This suggests that in analyzing relative
merit based on barnstars, the sign of the difference --- i.e. the simple
contrast between whether the voter has more barnstars than the candidate or
fewer --- is more salient that the actual numerical value of the difference.

The non-monotonicity around $0$, and the fact that it shows up so significantly
in both curves, suggests some intriguing conjectures about relative merit. In
particular, it suggests that voters are particularly critical of candidates
whose level of achievement is comparable to their own --- a contrast with the
simpler (and incorrect) hypothesis that the support of voters for candidates
should be purely monotonic in this relative level of achievement. Such a
conjecture forms an interesting connection to the recent lines of research in
social networks mentioned earlier, studying the roles played by relative
assessments in comparison to a peer group.

\begin{figure}[t]
    \centering
    \begin{tabular}{c}
    \includegraphics[width=0.4\textwidth]{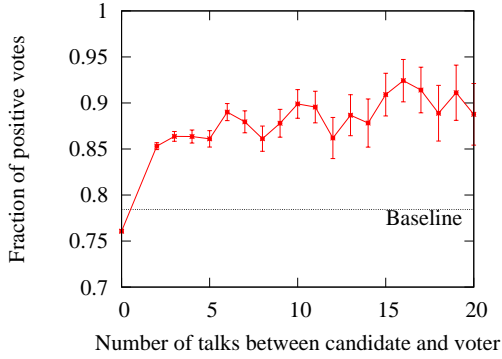} \\
    {\footnotesize (a) Positive vote fraction based on talk activity} \\
    \includegraphics[width=0.4\textwidth]{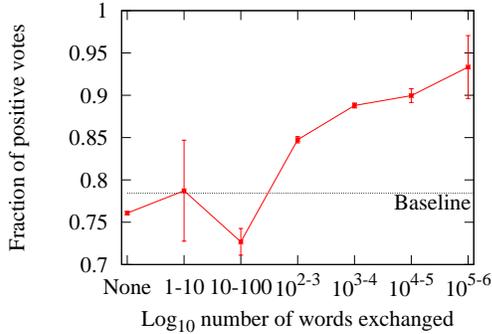} \\
    (b) {\footnotesize Positive vote fraction based on number of exchanged words}
    \end{tabular}
    \spaceB
    \caption{
Probability of $V$ voting positively on $C$ given the number of 
talk interactions and the total number of exchanged words between 
$V$ and $C$.  
}
    \label{fig:edits2}
    \spaceA
\end{figure}

\hdr{Direct Voter-Candidate Interactions.} Finally, we consider an even more
direct kind of relationship between a candidate $C$ and a voter $V$: the extent
to which $C$ and $V$ communicated prior to the election. We use edits that $C$
and $V$ made to each other's user-talk pages on Wikipedia as the trace data for
the history of communication between them.

Figure~\ref{fig:edits2} examines this by plotting the probability of a support
vote versus the number of talk-page edits between the candidate and the voter
(Figure~\ref{fig:edits2}(a)) and the total number of words exchanged by the
voter and the candidate on their respective talk pages
(Figure~\ref{fig:edits2}(b)). We see that there is a clear upward effect in
which the probability that $V$ will vote positively on $C$ tends to increase
with the amount of direct communication that the two have had.
Figure~\ref{fig:edits2}(a) in particular indicates that the simple existence or
non-existence of prior communication between $C$ and $V$ has a large effect on
the probability of a positive vote.

\section{Thresholds and Diversity in Voter Behavior}
\label{sec:thresh}
So far we examined how voters make decisions by comparing the candidate to
themselves. Now, we examine how voters evaluate the candidate in the context of
previous votes in the election. We explore how voters make decisions in the
context of a specific election, as it unfolds over time and in public. In this
context we are interested in threshold-based models that characterize changes
in voter behavior based on the current state of the election. Our investigation
addresses two basic issues: the relevance of threshold-based models, and the
diversity of thresholds across different voters.

\hdr{Threshold-Based Analysis of Voting Behavior.} The first issue is the
relevance of threshold-based models in analyzing how voters behave in an
election. For any vote cast in any election, we can define its {\em positive
precedent} to be the fraction of positive votes in the election up to that
moment. (In other words, if a vote was cast in an election at a moment when the
current vote count was 16 in favor and 4 against, then the positive precedent
of that vote would be $16/(16 + 4) = 0.8$.) Now, we define a {\em response
function} $f(x)$ as follows: over all votes with a positive precedent of $x$,
we set $f(x)$ equal to the fraction that were positive.

\begin{figure}[t]
    \centering
    \includegraphics[width=0.4\textwidth]{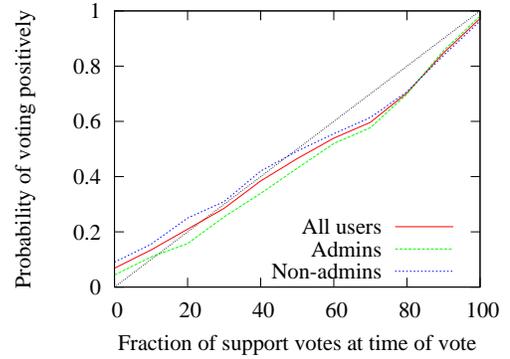}
    \spaceB
    \caption{Response function $f(x)$, and also aggregated separately
over the sub-populations of admins and non-admins.}
    \label{fig:global}
    \spaceA
\end{figure}

The fact that the elections are carried out sequentially in public forms the
motivation for this function: it is possible for a voter to know the current
fraction of positive votes (i.e. the positive precedent) at the moment she
casts her vote. If, for example, each voter flipped a coin with bias equal to
the current fraction of positive votes, and used this as her vote, then we
would see a response function $f(x) \approx x$. The extent to which a plot of
$f(x)$ deviates from the line $y = x$ can thus be taken as evidence
of a deviation from this baseline.

In Figure~\ref{fig:global}, we show a plot of the function $f(x)$, compared to
the diagonal line $y = x$. We see that $f(x) > x$ for small values of $x$ (up
to about $0.3$) and $f(x) < x$ for larger values of $x$
(above $0.3$, with the effect becoming particularly pronounced above $0.6$).
This is consistent with recent theories of sequential expressions of opinion in
on-line settings \cite{wu08opinion}; these theories argue that such deviations
represent a tendency for users to be more motivated to express an opinion when
it goes against the prevailing outcome. In this case, the argument would be
that users who view a candidate positively would be particularly motivated to
cast a positive vote (rather than simply not to vote at all) if they see that
the fraction of positive votes is particularly low. The corresponding reasoning
concerning negative opinions would support the observed downward deviation of
$f(x)$ at larger values of $x$.

Figure~\ref{fig:global} also shows plots of $f(x)$ aggregated over the
sub-populations of admins and non-admins. (Non-admins are
Unsucessful-candidates and Other users.) This partition of the full population
is a useful one in a number of our analyses in this section: although any
Wikipedia user is allowed to vote in a promotion election, the admins are the
ones who have successfully passed through the promotion process themselves, and
they are the ones most overtly charged with ensuring that Wikipedia functions
effectively. Thus, this division into the two sub-populations provides us with
a way to separately study the users who are most invested in the outcome of the
process and the users who are participating in the process but less invested.
In the plot, we find that when $x$ is small, $f(x) \approx x$ holds more
closely in the admin population than in the non-admin population; non-admins in
aggregate appear to be significantly more generous with their positive votes in
elections where the positive-vote fraction is low. At large values of $x$, the
two sub-populations agree very closely.

\begin{figure}[t]
    \centering
    \includegraphics[width=0.4\textwidth]{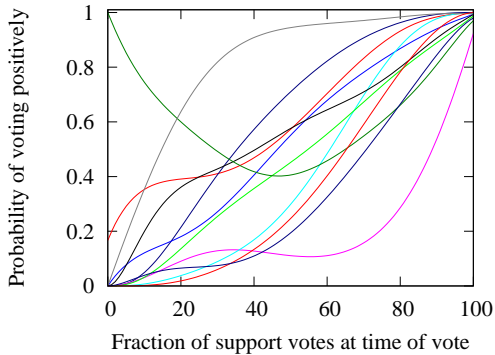}
    \spaceB
    \caption{Personal response functions for 11 users that voted on more than 400 elections.}
    \label{fig:threshold}
    \spaceA
\end{figure}

\begin{figure}[t]
    \centering
    \includegraphics[width=0.4\textwidth]{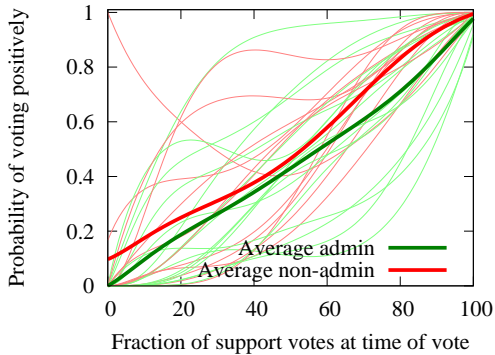}
    \spaceB
    \caption{Response functions of admins (green) vs. non-admins (red).
    Notice large variations in each sub-population.}
    \spaceA
    \label{fig:thresh2}
\end{figure}

\begin{figure*}[t]
    \centering
    \begin{tabular}{ccc}
    \includegraphics[width=0.32\textwidth]{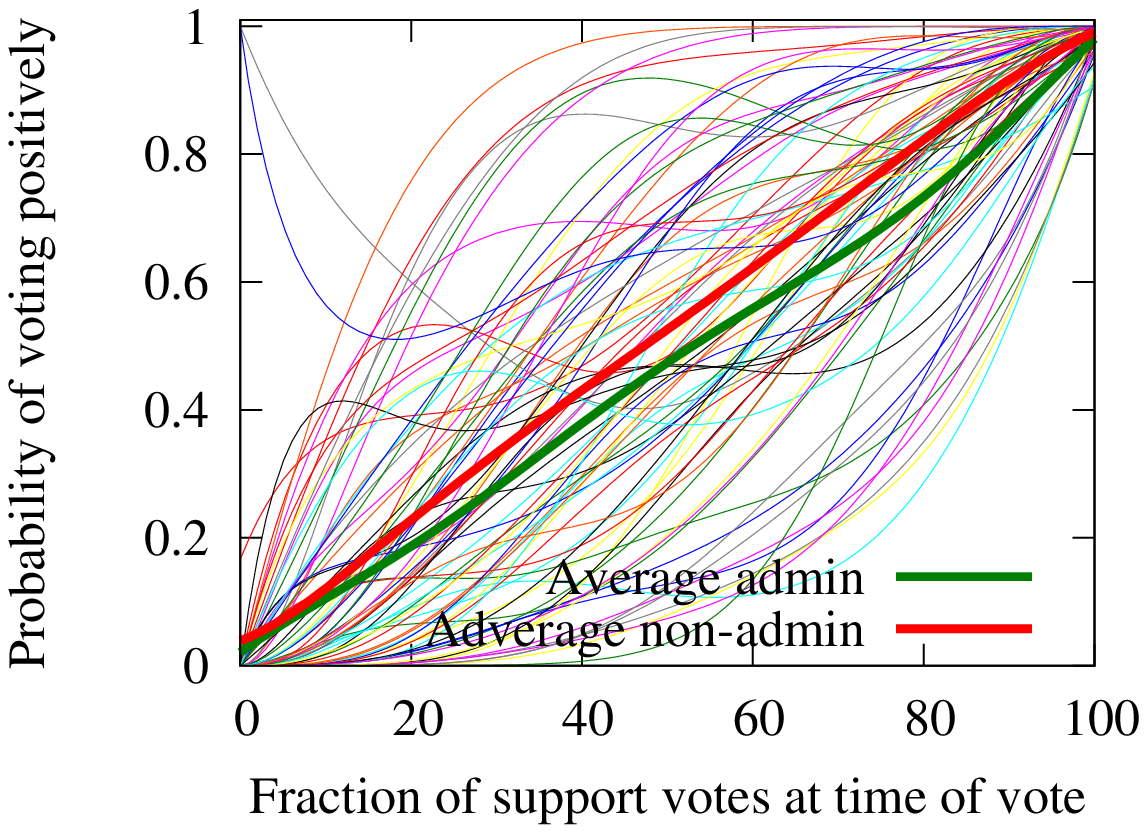} &
    \includegraphics[width=0.32\textwidth]{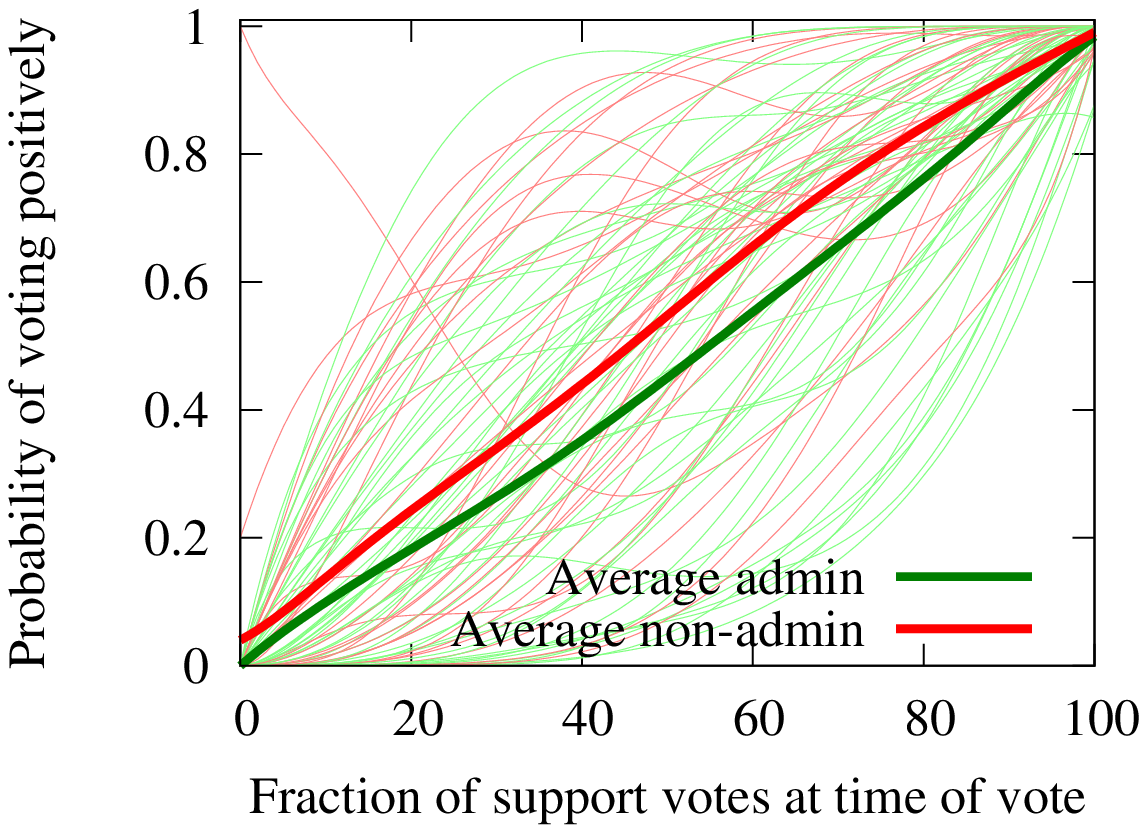} &
    \includegraphics[width=0.32\textwidth]{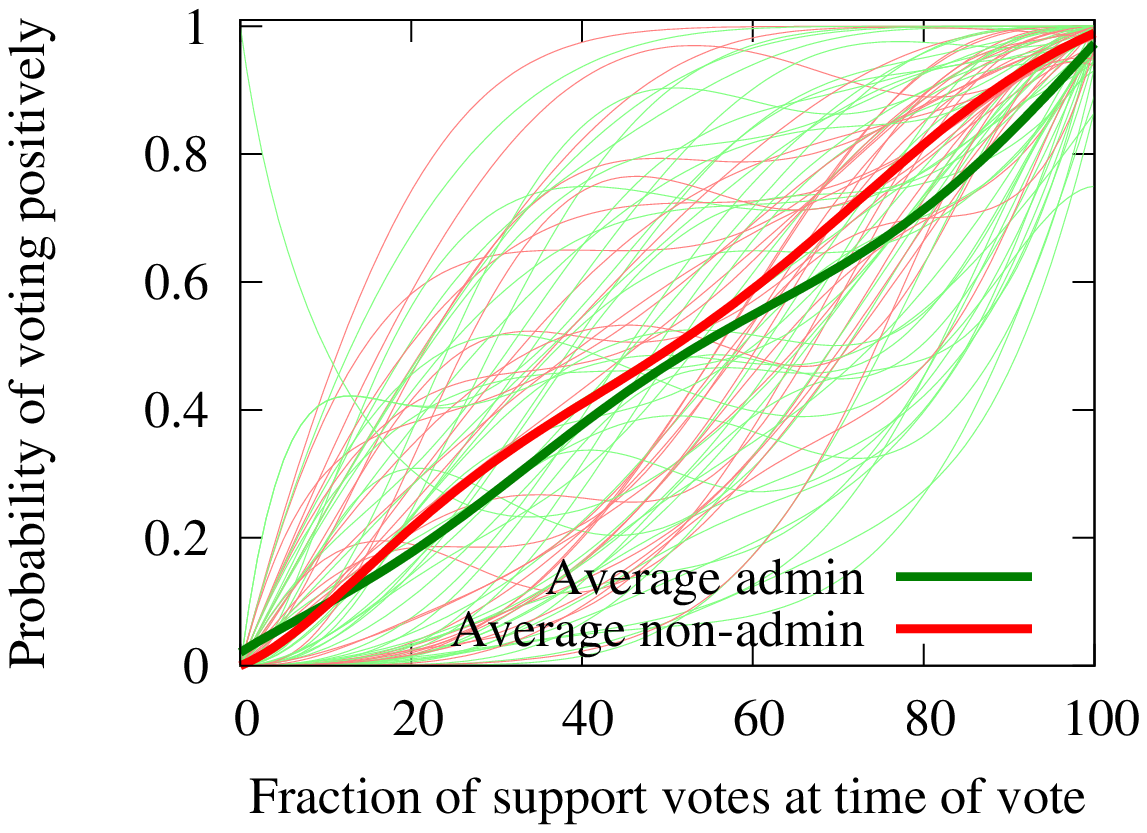} \\
     (a) Votes in all elections & (b) Votes in first half of elections & (c) Votes in second half of elections
    \end{tabular}
    \spaceB
    \caption{
    78 voters that participated in more than 200 elections. We take all their participation
    and only first and only second half of elections they participate in.}
    \label{fig:78voters}
    \spaceA
\end{figure*}

\hdr{Diversity of Individual Response Functions.} Just as we defined a function
$f(x)$ for the whole population, we can define a {\em personal response
function} $f_i(x)$ for each voter $i$. We define $f_i(x)$ to be simply the
analogue of $f(x)$ applied only the votes of voter $i$: over all votes with a
positive precedent of $x$ that were cast by $i$,
we set $f_i(x)$ equal to the fraction that were positive.

The natural worry in defining such a function is that there will not be enough
data on any individual $i$ to be able to meaningfully estimate $f_i(x)$. But on
Wikipedia, there are close to a hundred users who have voted in more than 200
elections, and for these users we can reasonably estimate $f_i(x)$ with
$x$ rounded to the nearest multiple of $0.1$.

Figure~\ref{fig:threshold} plots the estimated personal response functions
$f_i(x)$ for the 11 users who each voted in more than 400 elections, and
Figure~\ref{fig:thresh2} plots the estimated personal response functions for
the 28 users who each voted in more than 300 elections. What is immediately
striking is the considerable diversity in the shapes that these functions take.
Some users tend to vote overwhelmingly negatively whenever the current fraction
of positive votes is below $70\%$, while others are likely to vote positively
even when most votes thus far in the election have been negative. The
comparison of this diversity with the approximately diagonal shape of the
cumulative function $f(x)$ suggests that $f(x)$ represents in aggregate, over
the whole population, what is in reality an averaging of a highly diverse set of
individual response functions. This is an important issue to bear in mind
whenever we study such population aggregates; what is unusual in this case is
that we have a non-trivial collection of individuals with sufficiently
extensive personal histories in the system that we can actually build curves
for each of their individual patterns of behavior.

In Figure~\ref{fig:thresh2}, we also average separately over the admin and
non-admin sub-populations of this group of extremely frequent voters. The fact
that the admin curve is uniformly lower than the non-admin curve is consistent
with the more conservative approach to voting --- in aggregate --- that we saw
for the admin sub-population in Figure~\ref{fig:global} as well.

Finally, given the extensive personal histories of 78 users who have voted in
over 200 elections each, we can study how their voting behavior evolves over
time, by looking at a voter $i$'s personal response function $f_i^{(1)}(x)$
built only over the first half of the elections that $i$ participated in (in
chronological order), and the function $f_i^{(2)}(x)$ built only over the
second half of the elections that $i$ participated in. We find 
(in Figure~\ref{fig:78voters}) a general
tendency for voters to become more conservative in their use of positive votes
over time, and particularly for non-admins: the population average of
$f_i^{(2)}(x)$ over the non-admins in this group is clearly lower than the
population average of $f_i^{(1)}(x)$ over this group. For the admins, on the
other hand, the population averages of these two functions are more
similar, indicating a kind of aggregate stability in the voting behavior of
frequently voting admins as they ``grow older'' in Wikipedia.

\begin{figure}[t]
    \centering
    \begin{tabular}{cc}
    \includegraphics[width=0.22\textwidth]{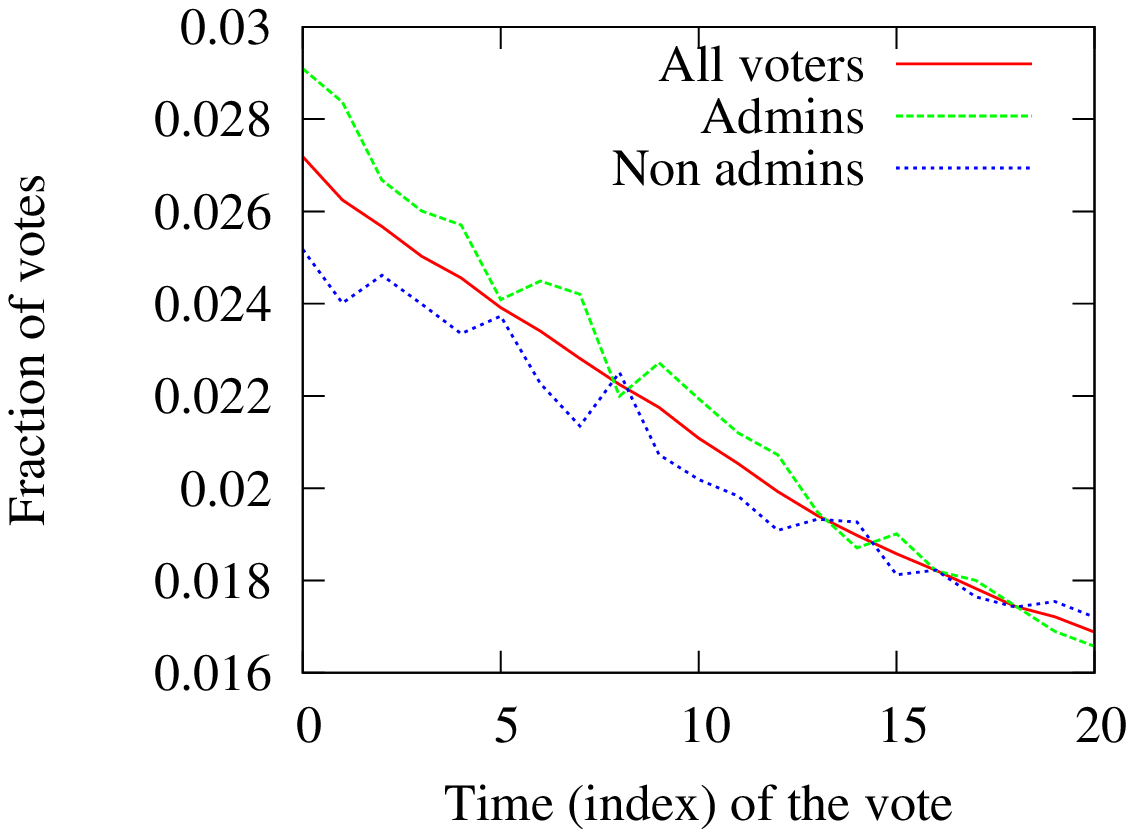} &
    \includegraphics[width=0.22\textwidth]{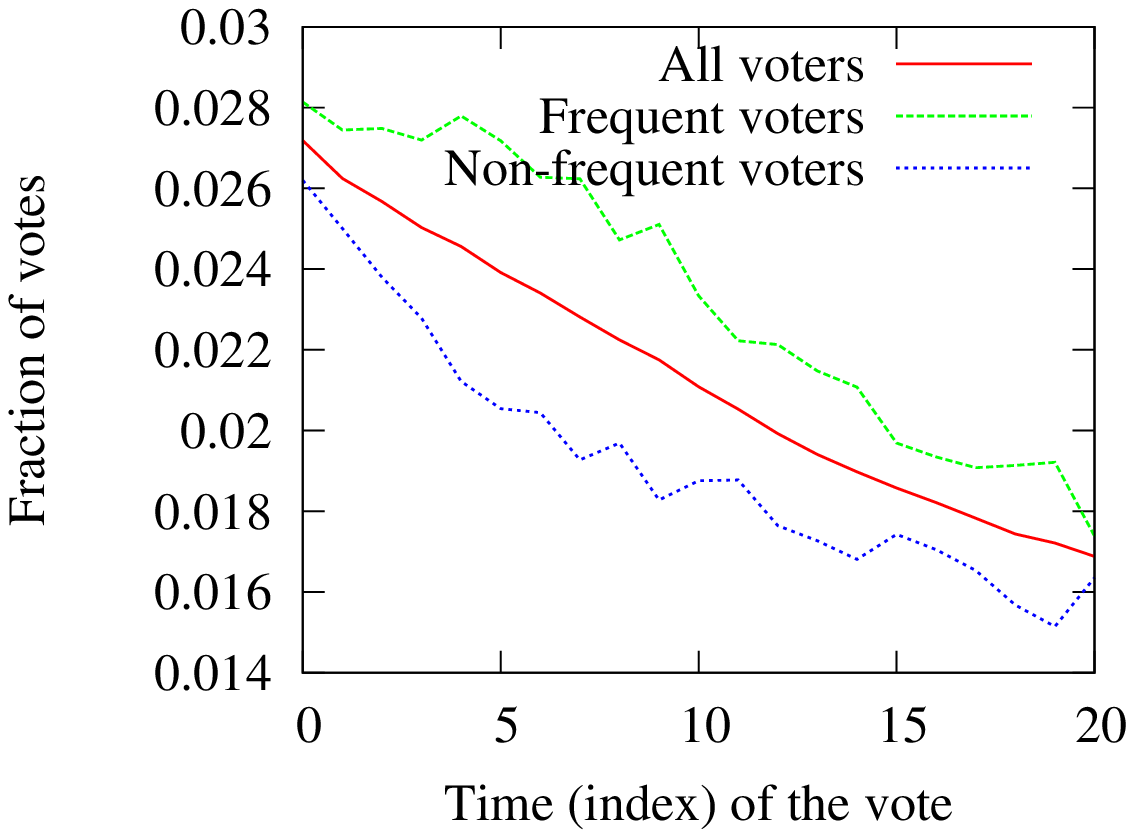} \\
     (a) Admins & (b) Frequent voters
    \end{tabular}
    \spaceB
    \caption{(a) Time when admins and non-admins cast their votes.
    (b) Time when heavy and non-heavy voters cast votes.
    }
    \label{fig:votetm}
    \spaceA
\end{figure}

\hdr{Timing of Entry.} A further issue is when, over the sequence of votes in
an election, different voters tend to arrive to cast their votes. For a given
voter, the timing of one's arrival affects how much information one has about
earlier votes, which may in turn affect one's own vote.

In Figure~\ref{fig:votetm}, we show when different sub-populations of voters
tend to cast their votes, relative to the overall population average. We find
much less difference between the admin and non-admin population (panel (a))
than we do between the populations of frequent and infrequent voters (panel
(b)); in this latter case, frequent voters tend to cast their votes earlier in
the election than non-frequent voters do. Interestingly, however, we do not
find any significant dependence between the time at which a voter casts her
vote and the positive/negative value of that vote.

\section{Dynamics of Elections over Time}
\label{sec:time}
In the previous section, we considered how the aggregate dynamics of an
election over time affected the decision of a particular voter at the time that
he or she arrived to cast a vote. We now examine the same process at much finer
resolution. We consider how the fine grained temporal dynamics of votes affects
the overall election outcome --- namely, we study how vote order affects the
election outcome. Our main finding is that the temporal order of votes ---
taking the final tally as given --- does not have a significant effect on the
outcome, with a few exceptions that we note below.

\begin{figure*}[!th]
    \centering
    \begin{tabular}{ccc}
    \includegraphics[width=0.21\textwidth]{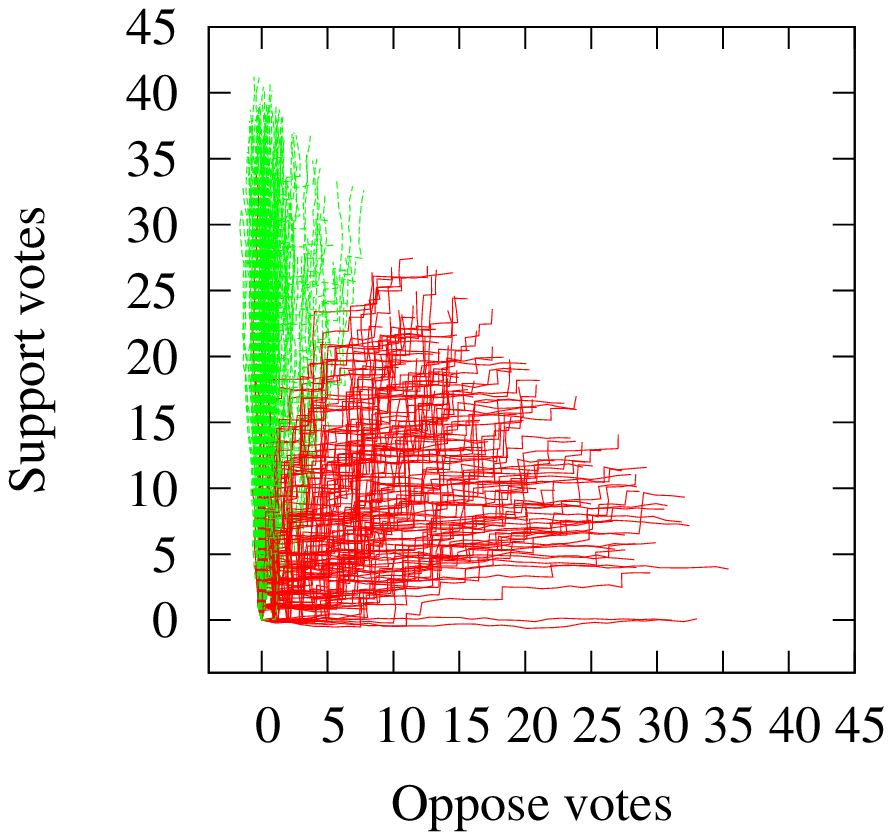}  &
    \includegraphics[width=0.28\textwidth]{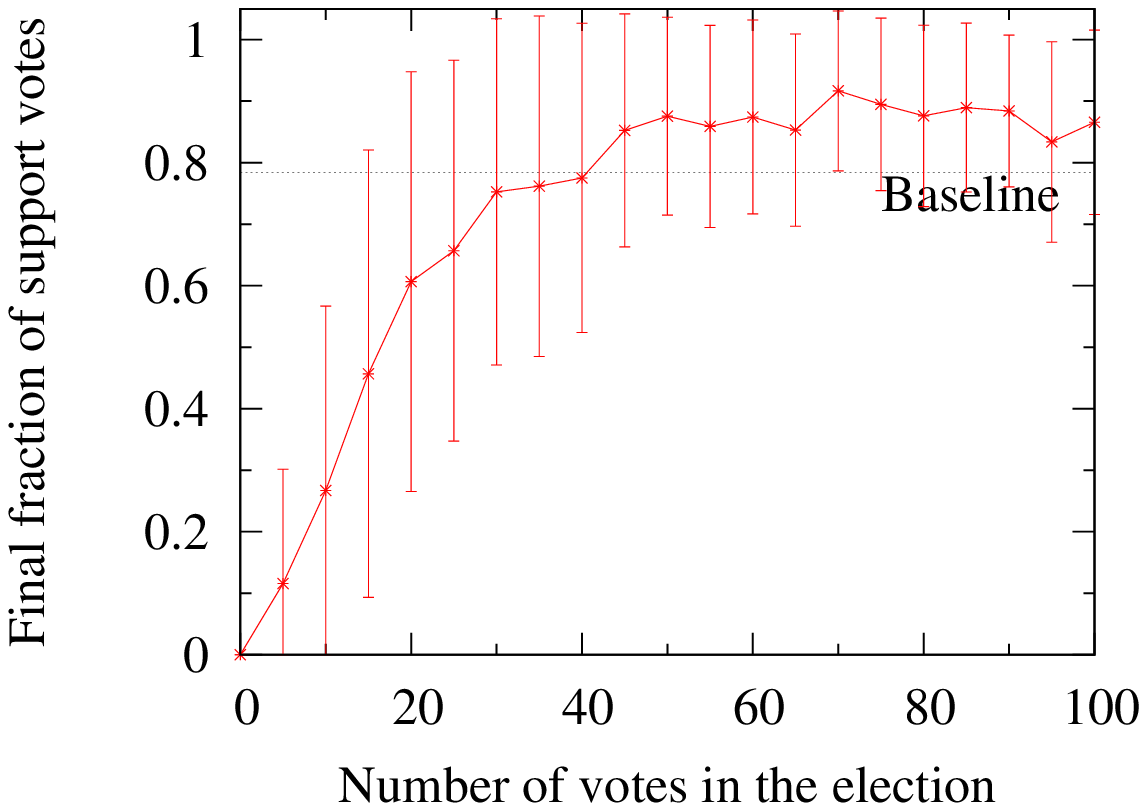} &
    \includegraphics[width=0.28\textwidth]{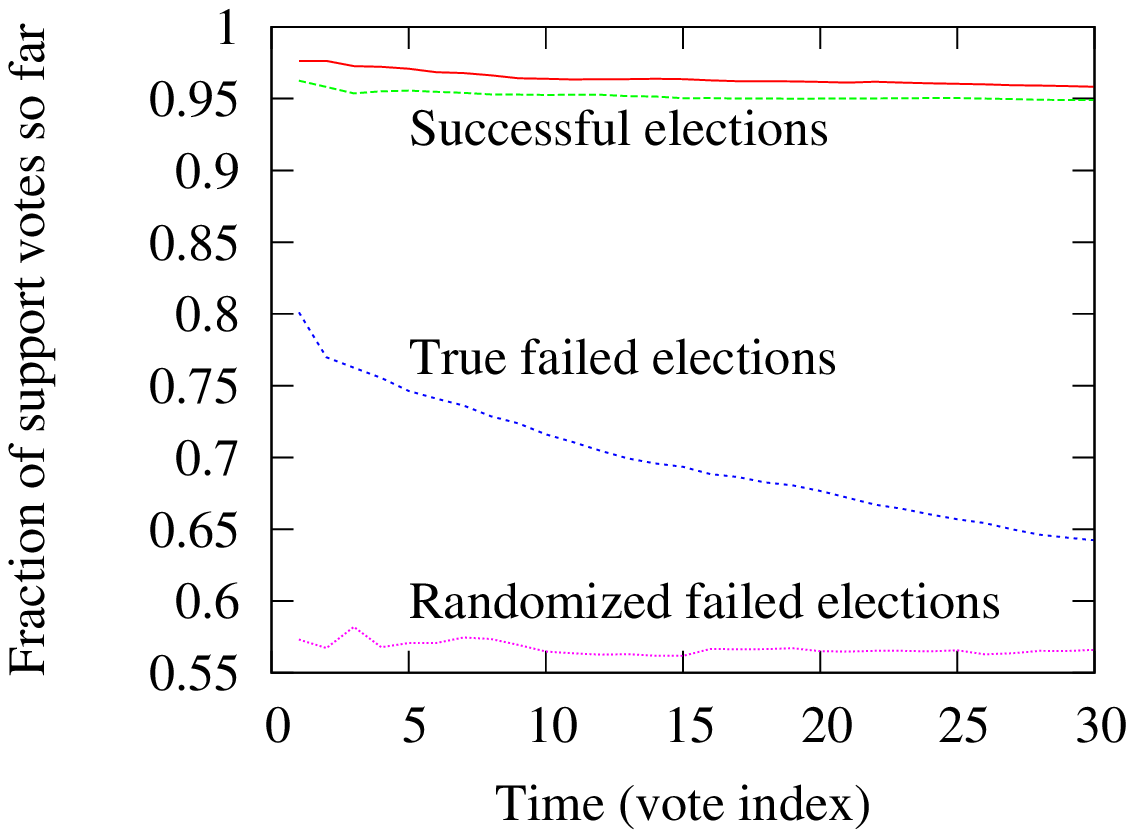} \\
    (a) Election progression & (b) Election success & (c) Election drift
    \end{tabular}
    \spaceB
    \caption{\footnotesize (a)  Progression of the
    election over time. Each election is a trail of support and oppose votes.
    (b) Probability that election results in promotion as a function of the number of votes.
    (c) Current fraction of positive votes as a function of time for failed and successful elections.
    Successful elections remain positive over time, while in negative elections the
    fraction of positive votes decreases over time.
}
    \label{fig:voteTime}
    \spaceA
\end{figure*}

\hdr{The History of an Election.} We make these question precise as follows.
Consider an election $e$, and let $p_e$ and $n_e$ be the total number of
positive and negative votes, respectively, that were cast in election $e$. The
full record of a public election with sequential voting also includes the order
in which the votes were cast; we define $p_e(j)$ and $n_e(j)$ to be the number
of positive and negative votes in election $e$ up through the point at which
$j$ votes in total had been cast. We define the {\em history} of the election
$e$ to be the sequence of points $\{(n_e(j),p_e(j)) : j = 0, 1, 2, \ldots, {\rm
length(e)}\}$, and we define the {\em running fraction} of positive votes to be
the sequence of fractions $\{p_e(j)/j : j = 1, 2, 3, \ldots\}$.
Figure~\ref{fig:voteTime}(a) provides a visual representation of the histories
of all elections; each history is a sequence of two-dimensional
points leading from origin at $(0,0)$ to the final tally $(n_e,p_e)$.

We begin with some basic initial observations about the histories of the
elections in our dataset. First, as shown in Figure~\ref{fig:voteTime}(b), the
Wikipedia bureaucrats who regulate the election process tend to end very
negative elections early, so that elections with long histories (of about 40
votes or more) tend to be the more successful ones. Second, we find that
unsuccessful elections tend to be ``top-heavy'' with an overrepresentation of
positive votes early, and an overrepresentation of negative votes later.
Figure~\ref{fig:voteTime}(c) provides an analysis of this: in an unsuccessful
election, the running fraction of positive votes declines over time, down to
the randomized baseline one gets by randomly permuting the order of the votes.
Successful elections, on the other hand, exhibit a relatively stable running
fraction of positive votes; it resembles the running fraction one would get
even if the order of the votes were randomly permuted. One conjecture is that
in unsuccessful elections, a candidate's close supporters vote early, leading
to an elevated fraction of positive votes, but this then declines as a broader
set of voters arrives.

\begin{figure*}[t]
    \centering
    \includegraphics[width=0.94\textwidth]{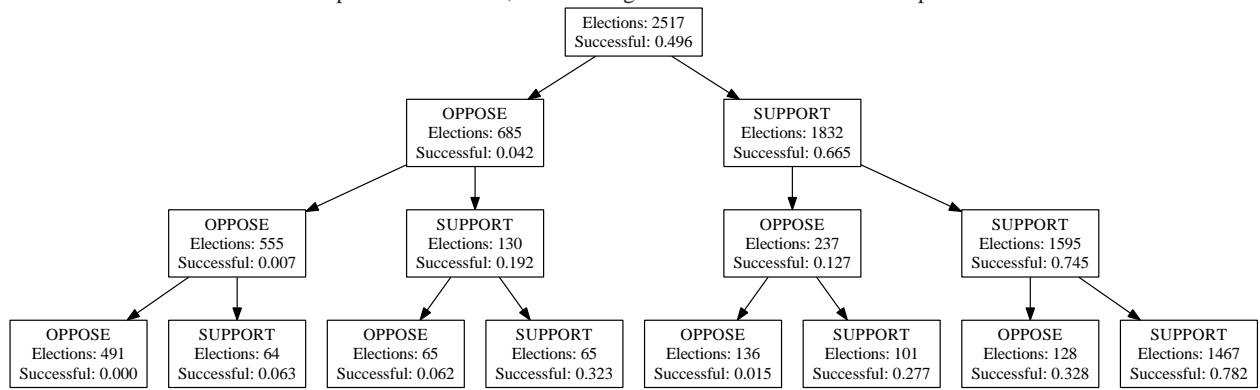}
    \spaceB
    \caption{Election tree. Notice that the order is not important but only
    the number of positive/negative votes.}
    \label{fig:tree}
    \spaceA
\end{figure*}

\hdr{The Prefix of an Election's History.} We also consider the specific effect
of the very first few votes in an election's history. Figure~\ref{fig:tree}
shows a tree of all possible prefixes of lengths 1, 2, 3 that an election's
history can have, and gives the total number of elections and the number of
successful elections for each such prefix.  Considering the prefixes that each
contain the same number of positive and negative votes reveals an interesting
pattern.

When considering the length-2 prefixes with a 1:1 tally we see that
(support,oppose) produces elections with a considerably lower rate of success,
0.127, than for (oppose,support), which has a success rate of of 0.192, even
though the vote count is the same at the end of these two prefixes. For
length-3 prefixes, there are three patterns with the tally 2:1 and three
patterns with the tally 1:2.  In each of these cases, the two patterns that do
not start with (support,oppose) have essentially the same success rates as one
another whereas the pattern that starts with (support,oppose) has a lower
success rate.  For example, the prefixes (support,support,oppose) and
(oppose,support,support) have rates of 0.328 and 0.323 respectively, whereas
(support,oppose,support) has a considerably lower rate of 0.277.

The discrepancy between these cases can be taken as a further reflection of the
idea from Figure~\ref{fig:voteTime}(c), that in unsuccessful elections, a
candidate's close supporters tend to vote early. This discrepancy also forms an
interesting contrast with results in economic theory suggesting that initial
positive votes can induce ``herding,'' elevating the probability of success
\cite{banerjee-herding,bikhchandani-fads}. The difference between
(support,oppose) and (oppose,support) shows the opposite contrast in our case,
due to the selection effects of a candidate's endorsers voting early.

\begin{figure}[t]
    \centering
    \begin{tabular}{cc}
    \includegraphics[width=0.22\textwidth]{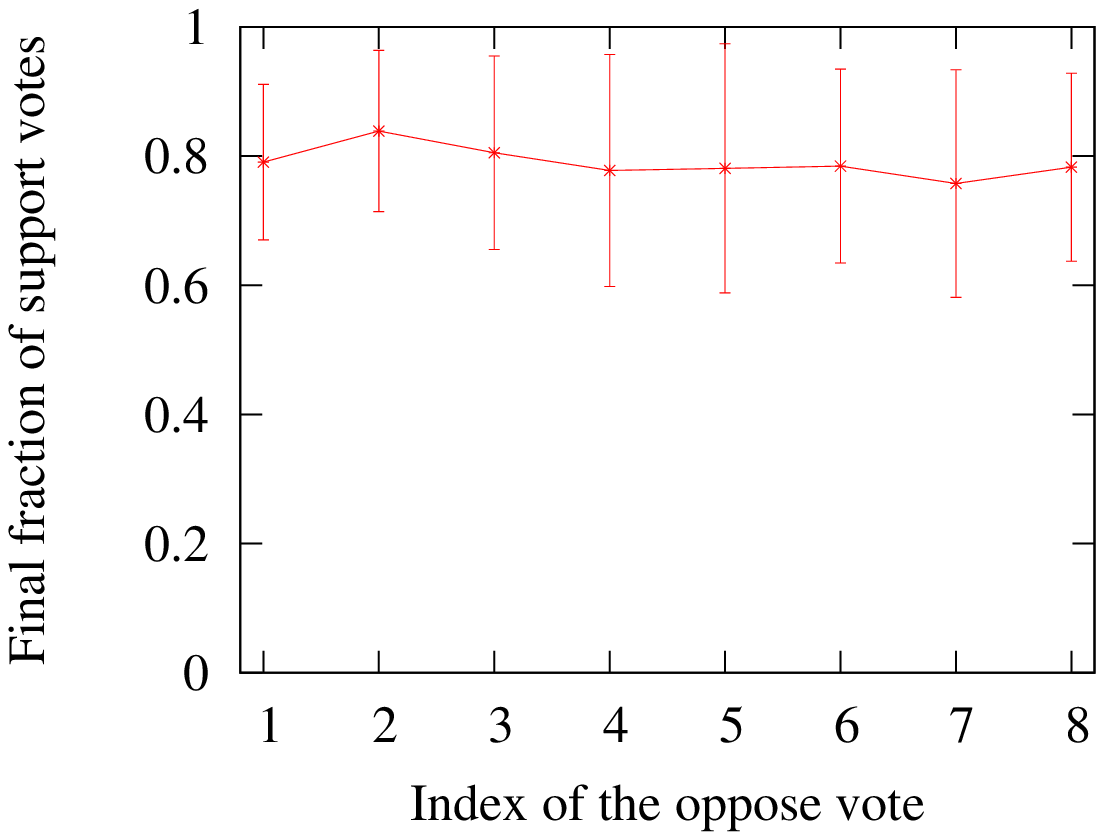} &
    \includegraphics[width=0.22\textwidth]{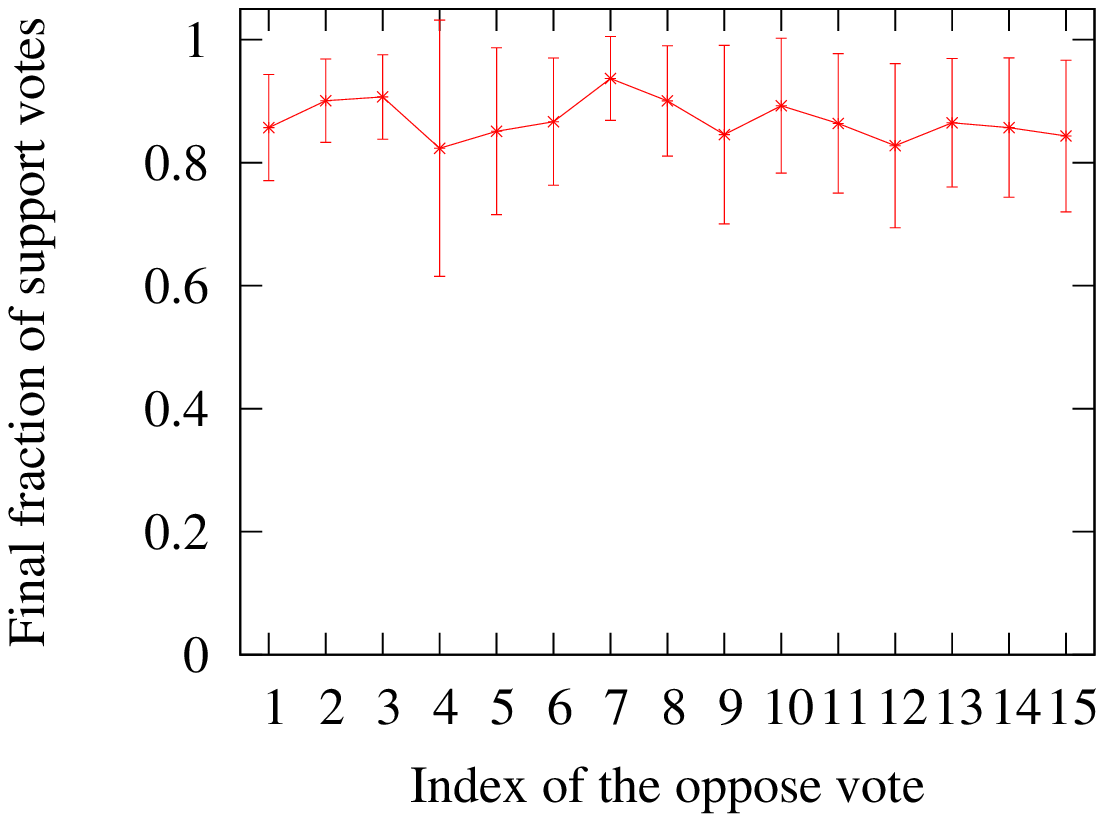} \\
    (a) First 8 votes & (b) First 15 votes
    \end{tabular}
    \spaceB
    \caption{Average fraction of support votes in the election as a function of the
    index of the first oppose vote.}
    \label{fig:oppVote}
    \spaceA
\end{figure}

Finally, we ask whether, in an election where support is very strong, the
timing of a single early negative vote can have a significant effect. As shown
in Figure~\ref{fig:oppVote}, it does not: in elections where the tally after
the first 8 votes is 7 positive and 1 negative (panel (a)), and in elections
where the tally after the first 15 votes is 14 positive and 1 negative (panel
(b)), the position in which the one negative vote occurs has essentially no
effect on the probability of a successful overall outcome. This again reflects
the ways in which timing effects appear to be more noticeable in unsuccessful
elections than successful ones.

\section{Conclusion}
\label{sec:conclusion}
As a case study of social-media governance,
we have investigated the Wikipedia promotion process from the perspective
of the voters engaged in group decision-making.
We have identified several forms of {\em relative assessment} that
play an important role in how voters make decisions; these include
how relative characteristics of voters and candidates
affect the probability of positive votes, as well as how
voters' decisions depend on the state of the election at the time
they cast their votes.
We have also investigated the temporal dynamics of the elections,
identifying ordering effects that contrast with standard theories
of herding and information cascades.

This style of analysis suggests a range of further interesting questions 
related to governance and deliberation.
It would be interesting to connect our findings on the relative
merit of voters and candidates more closely to the recent work of
Burt \shortcite{burt-neighbor-networks-book} and others on
the role that relative comparison plays in social networks.
We would also like to try integrating our analyses of
temporal dynamics in elections with Bayesian models of 
information cascades \cite{banerjee-herding}.
Finally, we believe that the style of analysis used here could
be productively combined with textual analysis of the content of
discussions that arise as part of deliberation on social-media sites;
such a hybrid of textual and structural approaches could well yield
further insights.

\bibliographystyle{aaai}

\begin{thebibliography}{}
\denselistA
\small
\bibitem[\protect\citeauthoryear{Backstrom \bgroup et al.\egroup
}{2006}]{backstrom-kdd06}
Backstrom, L.; Huttenlocher, D.; Kleinberg, J.; and Lan, X.
\newblock 2006.
\newblock Group formation in large social networks: Membership, growth, and
evolution.
\newblock In {\em KDD}, 44-54.

\bibitem[\protect\citeauthoryear{Banerjee}{1992}]{banerjee-herding}
Banerjee, A.
\newblock 1992.
\newblock A simple model of herd behavior.
\newblock {\em Quarterly Journal of Economics} 107:797--817.

\bibitem[\protect\citeauthoryear{Beschastnikh \bgroup et al.\egroup}{2008}]{beschastnikh09wiki}
Beschastnikh, I.; Kriplean, T.; McDonald, D.
\newblock 2008.
\newblock Wikipedian self-governance in action: Motivating the policy lens.
\newblock {\em ICWSM}. 

\bibitem[\protect\citeauthoryear{Bikhchandani \bgroup et al.\egroup}{1992}]{bikhchandani-fads}
Bikhchandani, S.; Hirshleifer, D.; and Welch, I.
\newblock 1992.
\newblock A theory of fads, fashion, custom and cultural change as information
cascades.
\newblock {\em J. Political Econ.} 100:992--1026.

\bibitem[\protect\citeauthoryear{Burke and Kraut}{2008}]{burke08promotion}
Burke, M., and Kraut, R.
\newblock 2008.
\newblock Mopping up: Modeling wikipedia promotion decisions.
\newblock In {\em Proc. CSCW '08},  27--36.

\bibitem[\protect\citeauthoryear{Burt}{2009}]{burt-neighbor-networks-book}
Burt, R.~S.
\newblock 2009.
\newblock {\em Neighbor Networks: Competitive Advantage Local and Personal}.
\newblock Oxford University Press.

\bibitem[\protect\citeauthoryear{Cosley \bgroup et al.\egroup
}{2005}]{cosley-norms}
Cosley, D.; Frankowski, D.; Kiesler, S.~B.; Terveen, L.~G.; and Riedl, J.
\newblock 2005.
\newblock How oversight improves member-maintained communities.
\newblock In {\em Proc. ACM CHI}, 11--20.

\bibitem[\protect\citeauthoryear{Danescu-Niculescu-Mizil \bgroup et al.\egroup
}{2009}]{mizil09opinions}
Danescu-Niculescu-Mizil, C.; Kossinets, G.; Kleinberg, J.; Lee, L.
\newblock 2009.
\newblock How opinions are received by online communities: A case study on
Amazon.com helpfulness votes.
\newblock In {\em WWW},  141--150.

\bibitem[\protect\citeauthoryear{Kossinets and Watts}{2006}]{kossinets-email}
Kossinets, G., and Watts, D.
\newblock 2006.
\newblock Empirical analysis of an evolving social network.
\newblock {\em Science} 311:88--90.

\bibitem[\protect\citeauthoryear{Kriplean \bgroup et al.\egroup}{2008}]{kriplean08barnstars}
Kriplean, T.; Beschastnikh, I.; and McDonald, D.~W.
\newblock 2008.
\newblock Articulations of wikiwork: uncovering valued work in wikipedia
through barnstars.
\newblock In {\em CSCW},  47--56.

\bibitem[\protect\citeauthoryear{Lerman and Galstyan}{2008}]{lerman08digg}
Lerman, K., and Galstyan, A.
\newblock 2008.
\newblock Analysis of social voting patterns on Digg.
\newblock In {\em 1st Workshop on Online Social Networks}.

\bibitem[\protect\citeauthoryear{Leskovec \bgroup et al.\egroup}{2006}]{leskovec-ec06}
Leskovec, J.; Adamic, L.; and Huberman, B.
\newblock 2006.
\newblock The dynamics of viral marketing.
\newblock In {\em Proc. ACM EC}.

\bibitem[\protect\citeauthoryear{Leskovec \bgroup et al.\egroup}{2010}]{leskovec-chi10}
Leskovec, J.; Huttenlocher, D.; and Kleinberg, J.
\newblock 2010.
\newblock Signed networks in social media.
\newblock In {\em Proc. CHI}.

\bibitem[\protect\citeauthoryear{Wu and Huberman}{2008}]{wu08opinion}
Wu, F., and Huberman, B.~A.
\newblock 2008.
\newblock How public opinion forms.
\newblock In {\em Proc. WINE},  334--341.

\end{thebibliography}


\end{document}